\documentclass[twocolumn,showpacs,preprintnumbers,amsmath,amssymb,prb]{revtex4}

\usepackage{graphicx}
\usepackage{dcolumn}
\usepackage{bm}

\begin{document}

\title{Local density of states at polygonal boundaries of $d$-wave superconductors}

\author{C.~Iniotakis, S.~Graser, T.~Dahm, and N.~Schopohl}

\affiliation{Institut f\"ur Theoretische Physik, Universit\"at T\"ubingen,\\
             Auf der Morgenstelle 14, D-72076 T\"ubingen, Germany}

\date{\today}

\begin{abstract}
Besides the well-known existence of Andreev bound states, the zero-energy
local density of states at the boundary of a $d$-wave superconductor 
strongly depends on the boundary geometry itself. 
In this work, we examine the influence of both a simple wedge-shaped boundary geometry 
and a more complicated polygonal or faceted boundary structure on 
the local density of states. 
For a wedge-shaped boundary geometry, we find oscillations of the zero-energy density of
states in the corner of the wedge, depending on the opening angle of the wedge.
Furthermore, we study the influence of a single Abrikosov vortex 
situated near a boundary, which is of either macroscopic or microscopic roughness.  
\end{abstract}


\maketitle

\section{Introduction}

The local density of states at the boundary of a superconductor is
a crucial factor in many experiments, for example tunneling measurements.
For conventional $s$-wave superconductors, the local density of states
at an insulating boundary is practically the same as in the bulk. In particular,
the specific boundary geometry is irrelevant. In the case of $d$-wave
symmetry, however, the situation is completely different. Due to Andreev
bound states \cite{Hu, Tanaka, Buchholtz}, a drastic enhancement of the low-energy density of
states can be observed at a straight flat surface appearing as a pronounced zero-bias
conductance peak \cite{Iguchi, Lesueur, Covington}. This effect
is maximal if the $d$-wave nodal direction is perpendicular to
the boundary and shrinks when the orientation is changed \cite{Tanaka, Iguchi}. For an angle
of 45 degrees between nodal direction and boundary, the Andreev bound states
disappear completely. Besides this well-known effect, it is important
to realize that for $d$-wave symmetry also the boundary geometry
itself can have strong influence on the local density of states. In
this work we examine the local density of states at the surface of
a $d$-wave superconductor for some basic examples of polygonal boundary
geometries and show that Andreev bound states are sensitive to the
boundary geometry. We also consider the additional influence of a single
Abrikosov vortex pinned near the boundary.

In a previous work we have shown, that the presence of an Abrikosov vortex
in the vicinity of the boundary of a $d$-wave superconductor has a
drastic effect on the zero-energy Andreev bound states. This effect appears as a 
surprisingly large shadow-like suppression of the zero-energy density of states between
vortex and boundary \cite{ShadowPRL}. The suppression is due to the
supercurrent flowing around the vortex, which locally shifts the quasiparticle
energies and leads to a splitting of the zero-energy peak. A similar 
splitting has already been studied for a homogeneous surface current\cite{Fogelstroem,CovingtonAprili,Aprili,Dagan}.
One of the purposes of the present study is to show that this vortex shadow
effect is stable and robust even if the boundary of the superconductor
possesses some roughness or faceting.

The paper is divided into five sections. In the next section, we
give a short introduction to the theoretical framework used for the
calculations. In the third section, we present some zero-energy local
densities of states for simple wedge-shaped boundary geometries and discuss
some elementary effects. After
that, we turn to the aspect of boundary roughness in section four, 
where we consider more complicated polygonal boundary structures. 
Finally, we end with our conclusions in section
five. 

\section{Theoretical background}
\subsection{The Riccati formalism}

In the limit  $k_F \xi \gg 1$, the equilibrium properties of a superconductor 
are contained in the Eilenberger propagator $\hat{g}(\vec{r},\vec{k}_F,i\epsilon_n)$. 
The propagator itself can be calculated by solving a transport-like equation first 
introduced by 
Eilenberger \cite{Eilenberger} and independently by Larkin and Ovchinnikov \cite{Larkin}. 
Here, we will use the so-called Riccati parametrization of the Eilenberger theory, which
was introduced in Ref.~\onlinecite{SchopohlMaki} and has proven to be very useful
for the numerical computation of the propagator 
(for more details see Ref.~\onlinecite{DGIS}).
In order to obtain a solution for a given Fermi vector
$\vec{k}_F$ at the point $\vec{r}_0$, the coupled Eilenberger equations are parametrized
along a trajectory passing through $\vec{r}_0$ in direction of the Fermi velocity corresponding
to $\vec{k}_F$
\begin{equation}
\vec{r}(x) = \vec{r}_0 + x \; \vec{v}_F(\vec{k}_F) 
\end{equation}
Along this trajectory, the Eilenberger equations can be transformed
to two decoupled differential equations of the Riccati type for 
the scalar functions $a(x)$ and $b(x)$:\cite{SchopohlMaki}
\begin{eqnarray}
\hbar v_F \partial_x  a(x) + \left[ 2 \epsilon_n + \Delta(x)^* a(x) \right] a(x)
- \Delta(x) & = & 0 \nonumber \\
\hbar v_F \partial_x b(x) - \left[ 2 \epsilon_n + \Delta(x) b(x) \right] b(x)
+ \Delta(x)^* & = & 0 \label{Riccati}
\end{eqnarray}
Here, we have already removed the vector potential $\vec{A}(\vec{r})$ via a gauge transformation. 
This can be done for the 
case of a single pinned Abrikosov vortex in a high-$\kappa$ superconductor,
since we concentrate on lengthscales smaller than the penetration depth $\lambda$. 
Together with the starting values in the bulk
\begin{eqnarray}
 a(-\infty) = \frac{\Delta (-\infty)}{\epsilon_n + \sqrt{\epsilon_n^2 + |\Delta
 (-\infty)|^2}} \nonumber \\
 b(+\infty) = \frac{\Delta^* (+\infty)}{\epsilon_n + \sqrt{\epsilon_n^2 + |\Delta
 (+\infty)|^2}}
 \label{Bulkab}
\end{eqnarray}
the functions $a$ and $b$ can be integrated in an easy and numerically stable way along
the trajectory.

The Eilenberger propagator $\hat{g}$ is related to the functions $a$ and $b$ by
\begin{equation}
\hat{g}=\frac{-\pi i}{1+ a b} \left( \begin{array}{cc}
1- a b & 2 i a \\
-2 i b & -1 + a b
\end{array} \right)
\end{equation}
Assuming a cylindrical Fermi surface, $\vec{k}_F$ depends only on an polar
angle $\theta$
\begin{equation}
\vec{k}_F = k_F \left( \begin{array}{c}
\cos \theta \\
\sin \theta
\end{array} \right)
\end{equation}
and the local density of states, normalized to the normal state density of
states  at the Fermi  level $N_0$, is given by 
\begin{equation}
N(\vec{r}_0,E) = \int_0^{2\pi} \frac{d \theta}{2 \pi}  {\rm Re} \;
\left[ \frac{1 - a b}{1+ a b} \right]_{i \epsilon_n \rightarrow E + i \delta}
\end{equation}
Practically, this means that the integrand has to be evaluated for many angles $\theta$, each 
angle corresponding to one particular trajectory through $\vec{r}_0$. The energy $E$ is 
taken with respect to the Fermi level and $\delta$ can be regarded as an effective scattering 
parameter which is proportional to an inverse mean free path.

In the calculations presented below, the specific boundary geometry of the 
superconductor has to be taken into account. 
Since we assume that the transmission coefficient of all the boundaries is zero, 
the boundary conditions\cite{Zaitsev, RainerBuchholtz, Shelankov} 
simplify considerably. In our case, a trajectory which hits the boundary is simply reflected 
like a ray of light. 
Thus, for the calculations in this work, a ray tracing procedure has been used,
which allows for multiple specular reflections. The Riccati equations (\ref{Riccati})
then have to be solved on such a multiply reflected trajectory (see for example the
trajectories in Fig.~\ref{AlpGamN}b)).

\subsection{Model for the Gap function}

We consider gap functions $\Delta (\vec{r},\vec{k}_F)$ that can be separated into
\begin{equation}
  \Delta (\vec{r},\vec{k}_F) = \Delta_0 \chi(\vec{k}_F) \psi(\vec{r})
\end{equation}
Here, the symmetry function for the $d_{x^2-y^2}$-wave is   
$\chi(\theta)=\cos(2\theta)$. In the following we want to concentrate
on the part $\psi(\vec{r})$ of the gap function, which covers the
spatial dependence. It can be factorized into modulus and phase 
\begin{equation}
  \psi (\vec{r}) = f_p(\vec{r}) e^{i \phi(\vec{r})}
\end{equation}
where we call the modulus $f_p(\vec{r})$ the profile function.
In principle, both the profile function and the phase $\phi(\vec{r})$ 
should be calculated selfconsistently via the gap equation. 
For simplicity, we have taken the profile function to be constant.
This is a good approximation for low temperatures due to the Kramer-Pesch effect \cite{KramerPesch}. 
Even for higher temperatures, the changes in the quasiparticle spectra 
due to this simplification are rather of a quantitative
than a qualitative nature. Compared to the constant model profile
function, the relevant differences of a more realistic profile function 
most often reside in comparatively small intervals along the trajectories.
For the situation studied in the following, this occurs near the boundaries or 
close to the vortex center. In these regions, the local density of states is
somewhat smeared out or softened, if a more realistic profile
function is used. However, the main features to be presented below 
remain unaffected, as we have checked for specific situations.

In the absence of a vortex, the phase $\phi(\vec{r})$ of the gap function 
simply is zero. For the case of a pinned Abrikosov vortex in the system
considered here, as an 
excellent approximation, the gradient of the phase is parallel to the 
current $\vec{j}(\vec{r})$ 
\begin{equation}
\vec{\nabla} \phi(\vec{r}) \; || \; \vec{j}(\vec{r})
\label{gradandj}
\end{equation}
and $\phi(\vec{r})$ is given as a solution
of the Laplace equation 
\begin{equation}
(\partial^{2}_x+\partial^{2}_y) \phi(\vec{r})=0
\end{equation}
with a "phase source" of $2\pi$ at the vortex position. In 
other words, integrating the phase gradient along any closed 
path around the vortex position must add up to a total of $2\pi$. 
Additionally, $\phi(\vec{r})$ has to fulfill von-Neumann boundary 
conditions 
\begin{equation}
\partial_n \phi(\vec{r}) = 0 
\end{equation} 
since currents are not allowed to cross any of the boundaries. 
Here, $\vec{n}$ denotes the normal vector of the boundary under
consideration. 

To solve the Laplace equation in two dimensions, 
we use the method of conformal mapping. 
This analytical technique can be found in many mathematical textbooks,
and in some specific physical textbooks as well. Nevertheless, in
the following we give the main aspects of conformal mapping in short.
First, the original two-dimensional domain, where the solution of
the Laplace equation is sought, is regarded as a part of the complex
plane. This domain is also referred to as physical domain, and it
often includes a nontrivial boundary geometry with specific boundary
conditions. The main idea and task is to map this physical domain
to a simple "model" domain by using analytic functions. Because
of the latter, the Laplace equation itself stays invariant under this
mapping. Thus, if the solution of Laplace's equation with the correct
boundary conditions is known in the model domain, also the original
problem is solved. The solution in the physical domain can then be
achieved by simply mapping each point to the model domain with the 
mapping function and taking the corresponding value of the solution
there. 

In our case of a wedge-shaped physical domain, the upper complex half space serves as the model plane
and $w,w_V$ shall denote elements thereof. Then, the function $\Phi$,
which is given by the formula 
\begin{equation}
\textrm{e}^{i\Phi}=\frac{w-w_V}{\left|w-w_V\right|}\cdot \frac{w^{*}-w_V}{\left|w^{*}-w_V\right|}
\label{mirrormodel}
\end{equation}
is the analytical solution of Laplace's equation 
\begin{equation}
 \partial_w \partial_{w^*} \Phi =0
\end{equation} 
in all points $w$ of the upper half plane with $w\neq w_V$. 
Integrating the phase gradient around the singularity at $w=w_V$
results in a total phase difference of $2\pi$. Thus, $w_{V}$ can
be regarded as the position of the phase vortex.
Also, von-Neumann boundary conditions are satisfied, 
since the phase gradient is parallel to the boundary (the real axis).
Of course, formula~(\ref{mirrormodel}) is analogous to the well-known 
mirror image ansatz in electrostatics, the first factor referring to 
a vortex at position $w_V$, the second to a virtual antivortex at
the mirrored position $w_V^*$.

\begin{figure}[t]
  \begin{center}
    \includegraphics[width=0.7\columnwidth]{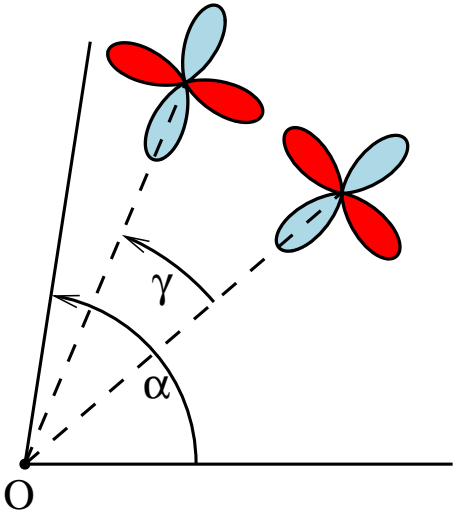}
   

    \caption{A $d_{x^2-y^2}$-wave superconductor with wedge-shaped 
    boundary geometry. $\alpha$ denotes the opening angle of the wedge. 
    The orientation between $d$-wave and wedge is parametrized by the
    angle $\gamma$, which gives the rotation of the full-gap direction
    with respect to the bisecting line of the wedge. The origin $O$ represents 
    the corner of the wedge.
    \label{Geometry}}
  \end{center}
\end{figure}

Since the correct solution of the problem in the model domain is
given by (\ref{mirrormodel}), the next and also last step is to 
find an analytical function which maps the physical domain to the 
upper half plane. 
It is well-known that the complex power function is analytic and that  
\begin{equation}
z(w)=w^{\frac{\alpha}{\pi}}
\end{equation}
maps the upper half plane to a wedge with opening angle $\alpha$. 
In particular, the positive real axis is mapped onto itself. 
Thus, for the physical domain being a wedge with opening angle $\alpha$
measured from the positive real axis, the sought mapping function
is just the inversion
\begin{equation}
w(z)=z^{\frac{\pi}{\alpha}}
\label{wmap1}
\end{equation} 
However, this mapping function is only valid for $\alpha\leq\pi$. 
For opening angles $\pi\leq\alpha\leq 2\pi$ one has to be careful 
about the branch cut discontinuity at the negative real axis. 
A possible mapping is then given by
\begin{equation}
w(z)=\textrm{e}^{i\pi(1-\frac{\pi}{\alpha})} \left[ 
z \textrm{e}^{i(\pi-\alpha)} \right]^\frac{\pi}{\alpha}
\label{wmap2}
\end{equation}
Finally, combining the solution (\ref{mirrormodel}) in the model plane 
with the above mapping functions $w(z)$, the phase factor of the 
gap function for a wedge-shaped boundary is given by
\begin{equation}
\textrm{e}^{i\phi(\vec{r})}=\frac{w(z)-w(z_V)}{\left|w(z)-w(z_V)\right|}\cdot \frac{w(z)^{*}-w(z_V)}{\left|w(z)^{*}-w(z_V)\right|}
\label{phasefactor}
\end{equation} 
On the right hand side, $z=r_x+i r_y$ is the position vector $\vec{r}$ in complex 
notation.
Analogous, the position of the vortex can be chosen with 
the parameter $z_V=r_{V,x}+i r_{V,y}$. 
The opening angle $\alpha$ of the wedge enters via the 
mapping functions (\ref{wmap1}) or (\ref{wmap2}), respectively.
In principle, the phase factor of more than one vortex is simply the product 
of several one-vortex solutions (\ref{phasefactor}). However, in the following 
we will confine ourselves to the case of only one single vortex.

\section{Wedge-shaped boundary geometry}
\begin{figure}[t]
  \begin{center}
    \includegraphics[width=0.95\columnwidth]{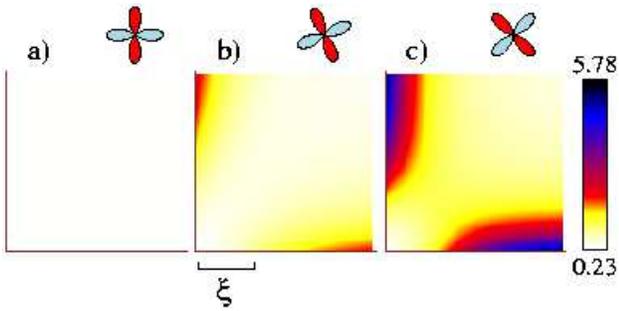}
   

    \caption{Zero-energy density of states of a $d_{x^2-y^2}$-wave 
    superconductor for a wedge-shaped boundary with opening angle $\alpha=\pi/2$. 
    The density of states is normalized to the density $N_0$ of the normal state. 
    The effective scattering parameter is chosen to be $\delta=0.1 \Delta_0$.
    The rotation of the $d$-wave with respect to the bisecting line is 
    a) $\gamma=-\pi/4$, b) $\gamma=-\pi/8$, c) $\gamma=0$.  
   \label{Keil90}}
  \end{center}
\end{figure}

\begin{figure}[t]
  \begin{center}
    \includegraphics[width=0.95\columnwidth]{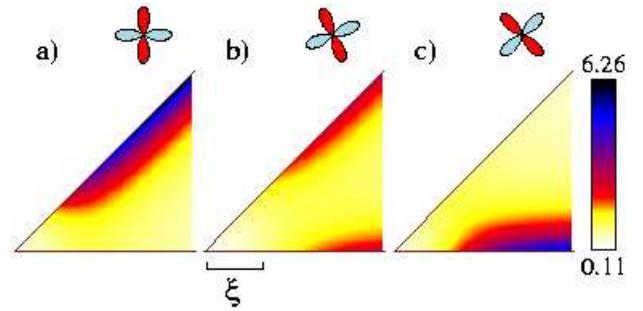}
   

    \caption{Zero-energy density of states of a $d_{x^2-y^2}$-wave 
    superconductor for a wedge-shaped boundary with opening angle $\alpha=\pi/4$. 
    The density of states is normalized to the density $N_0$ of the normal state. 
     The effective scattering parameter is chosen to be $\delta=0.1 \Delta_0$.
   The rotation of the $d$-wave with respect to the bisecting line is 
    a) $\gamma=-\pi/8$, b) $\gamma=0$, c) $\gamma=\pi/8$.  
    \label{Keil45}}
  \end{center}
\end{figure}

In this section, we investigate the zero-energy local density of states of 
$d_{x^2-y^2}$-wave superconductors with  
wedge-shaped boundary geometries. The situation is sketched in Fig. \ref{Geometry}. 
$\alpha$ denotes the opening angle of the wedge, $\gamma$ the angle between the 
bisecting line of the wedge and the direction of the maximum gap. 
Some examples for wedges with opening angles $\alpha=\pi/2$ and $\alpha=\pi/4$
can be seen in Figs. \ref{Keil90} and \ref{Keil45}. Far away from the corner,
the zero-energy density of states at the boundaries exhibit the
well-known Andreev bound states. If the nodal direction of the $d$-wave
is perpendicular to the specific boundary line, a maximum value is
reached (e.g. Fig. \ref{Keil90}c), both boundaries, or \ref{Keil45}a), upper boundary). 
For other orientations, the Andreev bound
states are smaller (Fig. \ref{Keil90}b) and \ref{Keil45}b)), or they even 
vanish if the direction of the maximum gap is perpendicular to 
the boundary (e.g. Fig. \ref{Keil90}a) or \ref{Keil45}a), lower boundary). 
In all the calculations presented here and in the following sections, 
the effective scattering parameter is set to $\delta=0.1\Delta_{0}$. 
This leads to a nonzero value of the zero-energy density of states in the bulk 
and limits the absolute value of the Andreev bound states at the boundaries. 

It is important to realize, however, that in a range from the corner,
which is given by the lengthscale $\xi=\hbar v_{F}/\Delta_{0}$,
the boundary geometry itself strongly influences the local density
of states. A very counterintuitive example can be seen in Fig. \ref{Keil90}c).
Although the zero-energy local density of states reaches its maximum
height along both boundary lines, the corner of the right-angled wedge
itself only exhibits the bulk value of $0.23$. For the smaller opening
angle $\alpha=\pi/4$, the local density of states in the tip of the 
wedge can even be lower than in the bulk: In Fig. \ref{Keil45}b), 
where $\gamma=0$, which refers to a maximum gap direction 
parallel to the bisecting line, the value is $0.11$. This is
barely larger than the bulk value corresponding to an
$s$-wave superconductor. 

\begin{figure}[t]
  \begin{center}
    \includegraphics[width=0.95\columnwidth]{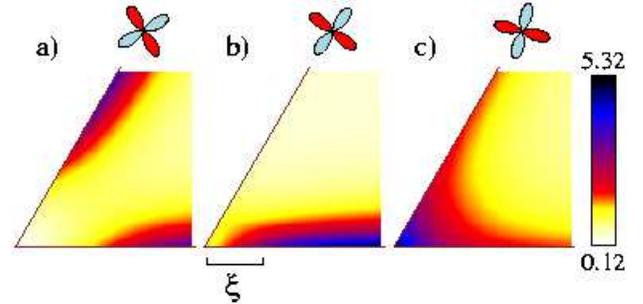}
   

    \caption{Zero-energy density of states of a $d_{x^2-y^2}$-wave 
    superconductor for a wedge-shaped boundary with opening angle $\alpha=\pi/3$. 
    The density of states is normalized to the density $N_0$ of the normal state. 
     The effective scattering parameter is chosen to be $\delta=0.1 \Delta_0$.
   The rotation of the $d$-wave with respect to the bisecting line is 
    a) $\gamma=0$, b) $\gamma=\pi/8$, c) $\gamma=\pi/4$.  
    \label{Keil60}}
  \end{center}
\end{figure}

By just taking a look at Figs. \ref{Keil90} and \ref{Keil45} one might
get the impression, that the zero-energy density of states in the corner
is always suppressed independently of the rotation angle $\gamma$ of the 
$d$-wave. However, this is not true in general. In Fig. \ref{Keil60} we
examine a wedge with opening angle $\alpha=\pi/3$. We can clearly see
that for the the nodal direction being parallel to the bisecting line 
($\gamma=\pi/4$), there is a strong increase of the zero-energy local 
density of states in the corner.

\subsection{Local density of states in the corner of the wedge}

In the following, we want to concentrate on the corner point $O$ and
systematically examine, why the local density of states seems to behave
so strange.

\begin{figure}[t]
  \begin{center}
    \includegraphics[width=0.95\columnwidth]{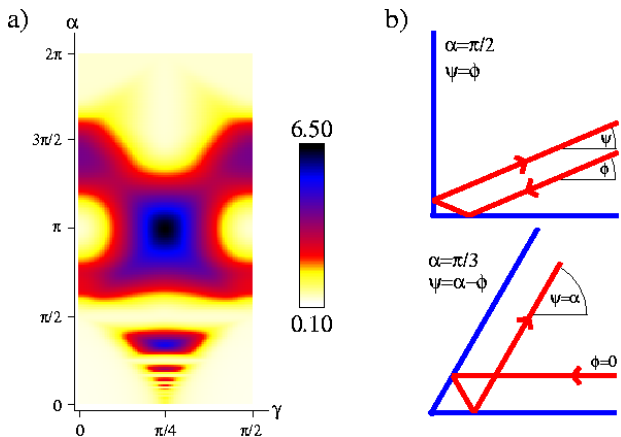}
   

    \caption{a) Zero-energy density of states in the corner $O$ of a $d_{x^2-y^2}$-wave 
    superconductor with wedge-shaped boundary geometry.
    $\alpha$ denotes the opening angle of the wedge, $\gamma$ the rotation of the
    maximum gap direction with respect to the bisecting line of the wedge. The global maximum
    in the middle of the picture corresponds to the well-known maximum of the 
    Andreev bound states at a straight boundary line, with the nodal direction of the $d$-wave
    being perpendicular to the boundary. b) Two examples for the relations (\ref{symmbez1})
    and (\ref{symmbez2}). The upper panel shows, that for an opening angle of $\alpha=\pi/2$
    an incoming trajectory always leaves the corner at the same angle. For an opening angle 
    of $\alpha=\pi/3$, the directions of the incoming and outgoing trajectories are 
    symmetric with respect to the bisecting line of the wedge (lower panel). 
    \label{AlpGamN}}
  \end{center}
\end{figure}

In Fig. \ref{AlpGamN}a), the zero-energy local density of states in the
corner is shown. The opening angle $\alpha$ of the wedge-shaped boundary
geometry corresponds to the vertical axis. The rotation angle $\gamma$ of
the $d$-wave is varied from $0$ to $\pi/2$ along the horizontal axis.
The absolute maximum value is right in the center of the figure. Here, 
$\alpha=\pi$ and $\gamma=\pi/4$ correspond to the well-known maximum 
Andreev bound states at a straight boundary line. 
The most interesting part of the picture can be seen at smaller opening
angles $\alpha\leq\pi$. If the nodal direction of the $d$-wave is kept 
fixed parallel to the bisecting line of the wedge (i.e. $\gamma=\pi/4$)
and the opening angle $\alpha$ is reduced, then the zero-energy 
local density of states clearly oscillates between high 
maxima and minima. A cut along this line showing the oscillations 
in the range $0<\alpha\leq\pi/2$ is shown in Fig. \ref{Oszi}.
Clearly, the minima of the oscillating zero-energy local density 
of states in the corner appear for opening angles $\alpha^{+}_{n}=\pi/(2n)$, 
while the maxima are very close to the angles $\alpha^{-}_{n}=\pi/(2n-1)$. 

\subsection{Explanation of the oscillations}

For opening angles $\alpha<\pi$, multiple reflections of a quasiparticle
trajectory occur, because the boundaries are hit several times.
Generally, for a given opening angle $\alpha$, the exact path of a quasiparticle
trajectory has to be determined by a raytracing algorithm. However,
there are some specific opening angles $\alpha_{n}=\pi/n$, which
have useful properties and are the key in understanding the oscillations.
We denote $\phi\in\left[0,\alpha\right]$ the angle of an incoming
trajectory before any reflection has occured, and call $\psi\in\left[0,\alpha\right]$
the angle of the final outgoing trajectory after all the multiple
reflections. Then, the specific opening angles $\alpha_{n}^{+}=\pi/(2n)$
have the property, that always
\begin{equation}
\psi=\phi
\label{symmbez1}
\end{equation}
while for $\alpha_{n}^{-}=\pi/(2n-1)$ we always have
\begin{equation}
\psi=\alpha-\phi
\label{symmbez2}
\end{equation}
In the first case, incoming and outgoing trajectories are parallel.
This means that the gap value is the same on the incoming and
outgoing part of the trajectory. In contrast,
in the second case the angles of incoming and outgoing trajectories
are symmetric with respect to the bisecting line of the wedge. 
For $\gamma=\pi/4$ (nodal direction along the bisecting line)
this means, that the gap possesses opposite sign on the
incoming and outgoing part of the trajectory, which results in large contributions to the 
zero-energy density of states in the corner.
Two typical examples for the different symmetry behaviour of 
in- and outgoing trajectory directions are shown in Fig.~\ref{AlpGamN}b).

\begin{figure}[t]
  \begin{center}
    \includegraphics[width=0.95\columnwidth]{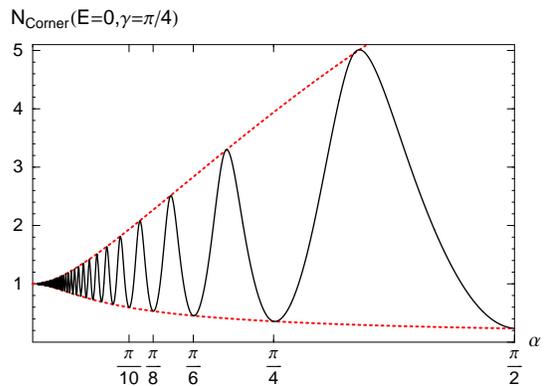}
   

    \caption{Oscillation of the zero-energy density of states in the corner 
    $O$ of a $d_{x^2-y^2}$-wave superconductor with wedge-shaped boundary geometry. 
    The nodal direction of the $d$-wave is fixed parallel to the bisecting line of
    the wedge, i.e. $\gamma=\pi/4$.  
    Decreasing the opening angle $\alpha$ leads to oscillations of the density
    of states in the corner, with minima appearing at $\alpha^{+}_{n}=\pi/(2n)$ and maxima
    close to $\alpha^{-}_{n}=\pi/(2n-1)$. Both upper and lower envelope (dotted lines) 
    are given analytically.
    \label{Oszi}}
  \end{center}
\end{figure}

Now we turn to the calculation of the local density of states in the
corner point $O$ of the wedge. For the trajectories passing through
this point, it is an excellent assumption to neglect all the complicated
details of the multiple reflections in the corner completely, since
they occur on a very small lengthscale along the trajectory path.
The only important things to keep are the relations between the angles
of incoming and finally outgoing trajectories. Thus, the effect of the wedge-shaped
boundary geometry is only a mixing of the bulk values (\ref{Bulkab}) of $a$ and
$b$ belonging to different angles. If we concentrate on the specific
opening angles $\alpha_{n}$ with the relations (\ref{symmbez1}) and (\ref{symmbez2}), 
it is possible to derive an analytic expression for the local density of 
states in the corner. 

The oscillations in Fig. \ref{Oszi} appear for the $d$-wave orientation
$\gamma=\pi/4$, where the nodal direction is parallel to the bisecting
line of the wedge. Based on relation (\ref{symmbez1}) for opening angles 
$\alpha_{n}^{+}=\pi/(2n)$, we find the lower envelope
\begin{eqnarray}
&&N_{\textrm{Corner}}^{+}(E=0,\gamma=\pi/4)  \\
&=&\frac{1}{\alpha}\int_{-\alpha/2}^{\alpha/2}d\phi\frac{1}{\sqrt{1+
\frac{\Delta_{0}^{2}}{\delta^{2}}\sin^{2} 2\phi}}
=\frac{1}{\alpha}F(\alpha,-\Delta_{0}^{2}/\delta^{2})\nonumber 
\end{eqnarray}
 Analogous, the upper envelope, which connects the values of the local
density of states for opening angles $\alpha_{n}^{-}=\pi/(2n-1)$,
is given by 
\begin{eqnarray}
&&N_{\textrm{Corner}}^{-}(E=0,\gamma=\pi/4) \\
&=&\frac{1}{\alpha}\int_{-\alpha/2}^{\alpha/2}d\phi\sqrt{1+\frac{\Delta_{0}^{2}}
{\delta^{2}}\sin^{2} 2\phi}=\frac{1}{\alpha}E(\alpha,-\Delta_{0}^{2}/\delta^{2}) \nonumber
\end{eqnarray}
Here, $F$ and $E$ are Elliptic Integrals of the First and Second
Kind. Both upper and lower envelope are shown as the dotted lines in
Fig.~\ref{Oszi}. As already mentioned before, we chose the effective scattering
parameter $\delta=0.1\Delta_{0}$. For the lower envelope, all the bulk
values stemming from an interval of length $\alpha$ around the nodal
line contribute to the local density of states in the
corner. Because of that, for $\alpha=\pi/2$ the whole quasiparticle 
spectrum in the corner is the same as in the bulk, since contributions
from the full $d$-wave are included. In particular, the zero-energy 
density of states in the corner is just the bulk value.
When the opening angle $\alpha$ is reduced, however, the result 
approaches the normal state value $1$. 
The upper envelope is dominated by bound
states, which occur by the mixing due to relation (\ref{symmbez2}), 
since the gap function changes sign at the bisecting line of the wedge. 
The highest contribution to the bound states is confined by the opening
angle $\alpha$. Thus, after the maximum for $\alpha=\pi$ at the
flat boundary, the upper envelope shrinks to the normal state value
$1$ for smaller opening angles.

If the maximum gap direction of the $d$-wave is parallel to the bisecting
line of the wedge, i.e. $\gamma=0$, the sign of the gap function
is symmetric with respect to the bisecting line. Then, the mixing
(\ref{symmbez2}) of the specific opening angles $\alpha_{n}^{-}$ generates
nothing but bulk contributions, too. Consequently, for all the opening
angles $\alpha_{n}=\pi/n$, no bound states appear at all. The corresponding
values of the zero-energy local densities of states in the corner
are on the same curve for both $\alpha_{n}^{+}$ and $\alpha_{n}^{-}$:
\begin{eqnarray}
&&N_{\textrm{Corner}}(E=0,\gamma=0)=
\frac{1}{\alpha}\int_{-\alpha/2}^{\alpha/2}d\phi\frac{1}{\sqrt{1+\frac{\Delta_{0}^{2}}{\delta^{2}}\cos^{2}2\phi}} 
\nonumber \\
&=&\frac{\frac{\delta}{\Delta_{0}}}{\sqrt{\frac{\delta^{2}}{\Delta_{0}^{2}}+1}}\frac{1}{\alpha}
F(\alpha,1/(1+\delta^{2}/\Delta_{0}^{2})) \label{Gamma0Linie}
\end{eqnarray}

For $\alpha_{1}=\pi$ and again for $\alpha_{2}=\pi/2$, we have the
same density of states both in the bulk and in the corner. For smaller
opening angles, only the contributions from the full-gap direction
remain. The local density of states in the corner approaches the shape
of a bulk $s$-wave spectrum. Because of that, the zero-energy weight
approaches the $s$-wave bulk value, as already pointed out before.
Although there are no zero-energy bound states in the corner of the
wedge for the specific opening angles $\alpha_{n}=\pi/n$, multiple 
reflections lead to considerable bound states for angles $\pi/2<\alpha<\pi$. 
For arbitrary opening angles $\alpha<\pi/2$, however, the deviation
of the zero-energy density of states in the corner from Eq.~(\ref{Gamma0Linie})
determined by the $\alpha_{n}$ is small.  

To summarize this section: Whether Andreev bound states appear in the corner of
a wedge or not depends on both opening angle $\alpha$ of the wedge  and orientation
$\gamma$ of the $d$-wave. For $\alpha<\pi/2$, Andreev bound states in the corner are suppressed
in most of the parameter space. However, near the specific opening angles
$\alpha_{n}^{-}=\pi/(2n-1)$ and for orientations about $\gamma=\pi/4$, Andreev bound states
get induced in the corner. Fig.~\ref{AlpGamN}a) provides a map, showing for which combinations 
of angles Andreev bound states appear.

\subsection{Wedge and Vortex}
We already discussed the influence of a single Abrikosov 
vortex, which is pinned near a straight smooth boundary, 
on the local quasiparticle spectrum in a previous work \cite{ShadowPRL}. 
The result is a suppression of the zero-energy Andreev bound states 
in a shadow-like region extending from the vortex to the boundary.
This effect is due to the flow field of the phase gradient around the vortex, 
which leads to a local shift of the quasiparticle energy along the trajectory.
As a consequence, the zero-energy Andreev bound states at the boundary are 
suppressed and the spectral weight is shifted towards higher quasiparticle
energies as discussed in Ref.~\onlinecite{ShadowPRL}.

\begin{figure}[t]
  \begin{center}
    \includegraphics[width=0.95\columnwidth]{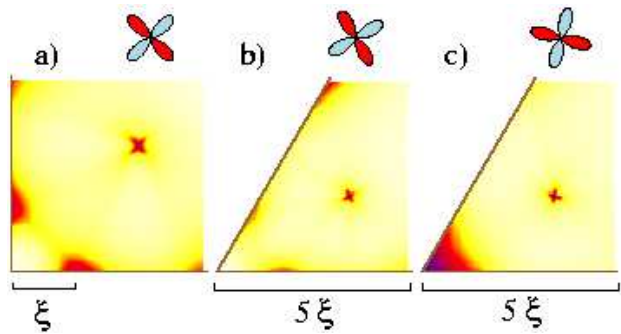}
   

    \caption{Zero-energy density of states of $d_{x^2-y^2}$-wave 
    superconductors with wedge-shaped boundary geometries. 
    A single Abrikosov vortex is pinned at a distance of 2 $\xi$ 
    from each boundary line. All pictures presented here 
    correspond to a specific wedge and orientation already shown 
    earlier without a vortex.  
    The same color palettes have been used for both the 
    corresponding plot without vortex and the plot given here.
    a) This picture refers to Fig. \ref{Keil90}c). 
    The vortex casts a shadow-like suppression on the Andreev 
    bound states at each boundary. 
    b) This picture refers to Fig. \ref{Keil60}a). Again, 
    the Andreev bound states at the boundaries are locally suppressed.
    c) This picture refers to Fig. \ref{Keil60}c). The bound states
    in the corner induced by the boundary geometry are also affected but 
    remain still considerable. Please note, that b) and c) are given in 
    a smaller scale than the corresponding pictures.  
    \label{KeilVortex}}
  \end{center}
\end{figure} 
 
Some examples of a single Abrikosov vortex, which is pinned near a wedge-shaped
boundary geometry, are shown in Fig. \ref{KeilVortex}. Here, we have calculated
the phase distribution around the vortex using the conformal mapping procedure
described in section II B. Each picture 
corresponds to a picture with the same wedge geometry and $d$-wave orientation
already presented in earlier figures. As a main effect, the existence of the 
vortex and its inhomogeneous phase lead to a local suppression of the zero-energy 
Andreev bound states at the boundaries. The center of the suppression seems
to be approximately that point of the boundary, which lies closest to the vortex.

If we concentrate on the very corner of the wedge, however, the characteristic
changes of the local density of states induced by the boundary geometry itself 
are only slightly affected by the presence of the vortex. Of course, the induced
bound states in the corner of Fig. \ref{Keil60}c), for example, get reduced 
by about 30 \% in Fig. \ref{KeilVortex}c) because of the vortex. 
Nevertheless, the zero-energy density of states is still strongly increased 
in the corner compared to the bound states at the boundary nearby.
An induced suppression of the zero-energy density of states in 
the corner is practically not affected by the vortex as can be seen in 
Figs. \ref{KeilVortex}a) and b).
It is important to realize, however, that the shown suppressions of the  
zero-energy local density of states are not of the same nature. On the one hand,
the suppression due to the vortex is the result of a splitting of the sharp zero-energy peak 
in the quasiparticle spectrum. On the other hand, the low value of the zero-energy density of
states because of the 
boundary geometry is just a natural consequence, when the quasiparticle spectrum in the corner
is given by a $d$- or even $s$-wave bulk spectrum.

\begin{figure}[t]
  \begin{center}
    \includegraphics[width=0.95\columnwidth]{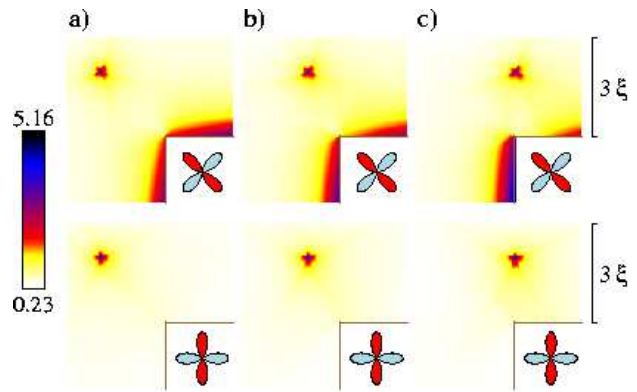}
   

    \caption{Zero-energy density of states of a $d_{x^2-y^2}$-wave 
    superconductor for a wedge-shaped boundary with opening angle $\alpha=3\pi/2$. 
   A single Abrikosov vortex is situated at a vertical distance of $2\xi$ from the 
   corner of the wedge. The horizontal distance of the vortex is a) $2\xi$, b) $\xi$, c) 0.
   In the top row, the nodal direction of the $d$-wave is perpendicular to both boundaries, 
   which corresponds to $\gamma=0$.
   Below, the nodal direction is parallel to the bisecting line, i.e. $\gamma=\pi/4$.  
    \label{StumpVortex}}
  \end{center}
\end{figure}

Some further examples of a wedge with opening angle $\alpha=3\pi/2$ and an Abrikosov 
vortex at different positions can be seen in Fig. \ref{StumpVortex}. In the top row,
the shadow effect can be observed again. Below, there are no bound states at the
boundaries at all, and the influence of the vortex rather leads to a very small increase
of the zero-energy density of states. 

\section{Polygonal boundary structure}

In this section, we want to study the influence of surface roughness and surface
faceting on the vortex shadow effect introduced in Ref.~\onlinecite{ShadowPRL} in
order to see how stable this effect is. There have been several different
suggestions in the literature how surface roughness can be implemented within
Eilenberger theory (see for example Refs.~\onlinecite{Shelankov,Fogelstroem,Thuneberg}).
Here, we focus on the zero-energy density of states
in the vicinity of a boundary line for two different models of either microscopic 
or macroscopic roughness with respect to the lengthscale of the coherence length.
We begin with a very simple model of microscopic roughness in the next subsection.
After that, we present results for a polygonal boundary structure, which can
be regarded as a macroscopically rough, faceted surface. In contrast to previous
work we take into account the complicated current redistribution along the
faceted surface using conformal mapping techniques.

\subsection{Microscopic roughness}

\begin{figure}[t]
  \begin{center}
    \includegraphics[width=0.95\columnwidth]{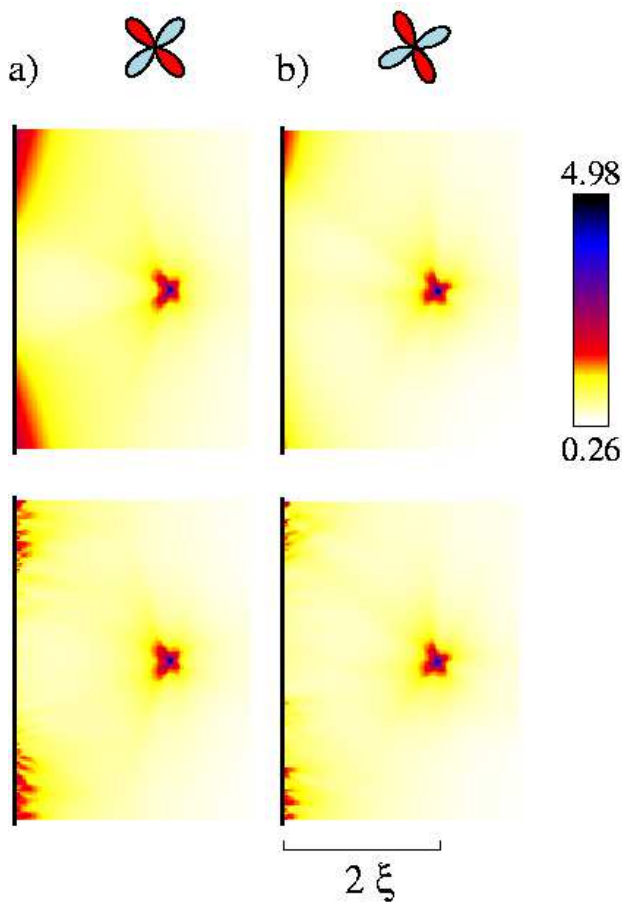}
   

    \caption{Zero-energy density of states of a $d_{x^2-y^2}$-wave 
    superconductor in the vicinity of a surface. This corresponds to
    an opening angle of the wedge of $\alpha=\pi$.
   A single Abrikosov vortex is located at a distance of $2\xi$ from the boundary. 
   In the top row, the boundary is smooth. In the lower row, the 
   boundary is microscopically rough.
   a) The nodal direction of the $d$-wave is perpendicular to the boundary, 
   which corresponds to $\gamma=\pi/4$.
   b) The rotation angle is $\gamma=\pi/6$. Compared to a), the orientation
   changed by 15 degrees.  
    \label{Specular}}
  \end{center}
\end{figure}

A simple model of a microscopically rough surface is given by an 
arrangement of randomly tilted tiny mirrors along the boundary line.
The size of each mirror
is taken to be about $\xi/40$, which is of an atomic lengthscale for high-$T_c$ 
superconductors. Each mirror is randomly tilted. We have chosen a Gaussian distribution 
between $-\pi/4$ and $+\pi/4$ for all the tilt angles, with a tilt angle of $0$ 
corresponding to a parallel alignment of mirror and boundary line.  
Additionally, we made the simplification, that the mirrors do not extent into the
superconductor. Their only function is to reflect an incoming trajectory according
to the mirror orientation, thus producing a kind of random scattering at the boundary.
There are some further consequences of the simplification made above. On the one hand,
the flow field inside the superconductor is not affected by the boundary roughness and
is simply the one of a wedge-shaped boundary geometry with opening angle $\alpha=\pi$.
On the other hand, multiple reflections at the boundary and the corresponding effects 
presented in the previous section are excluded. However, it is important to note, that 
for any boundary orientation, there exist zero-energy bound states in some regions 
of the boundary line because of the microscopic roughness. This is in agreement with earlier
calculations of the zero-bias conductance peak for a rough surface \cite{Fogelstroem}.   

How is the shadow effect due to a vortex affected by such a microscopically rough
surface? The zero-energy local
density of states for a boundary line with one specific randomly generated mirror arrangement
and an adjacent Abrikosov vortex is shown in Fig. \ref{Specular}. 
In column a), the nodal line of the $d$-wave is perpendicular to the boundary, 
which corresponds to a (110) surface or $\gamma=\pi/4$, respectively. 
Although the appearance of the Andreev bound states along the boundary line 
changes in a characteristic way because of the microscopic roughness, the suppression
of the Andreev bound states in a shadow-like region clearly persists.
In b), the orientation between boundary and $d$-wave is changed by $15$ degrees.
The suppression of the bound states remains, but it is obviously no more symmetric,
neither for the smooth nor the microscopically rough boundary.
Since the modulus of the phase gradient around the vortex is symmetric, 
this asymmetry cannot be explained
by a Doppler-shifted energy spectrum, where only local surface currents are taken into account.
In fact, generally the whole quasiparticle "history" along its trajectory is very important 
and may not be neglected. 
Only in the case of a homogeneous flow field \cite{Fogelstroem}, for example far away from the vortex, it becomes sufficient
to consider solely the local current. 
 
\subsection{Macroscopic roughness}

\begin{figure}[t]
  \begin{center}
    \includegraphics[width=0.88\columnwidth]{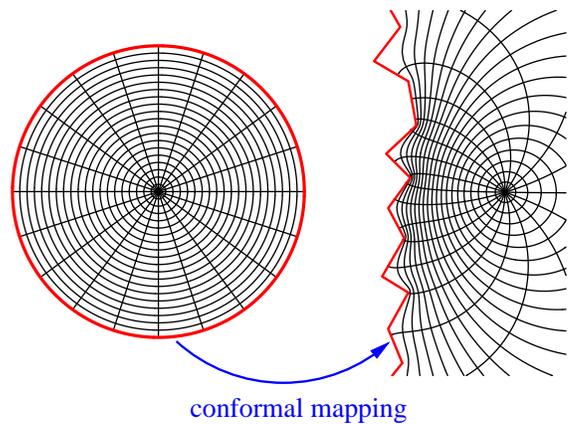}
  

    \caption{ 
  The conformal mapping procedure used to obtain the phase distribution
    around an Abrikosov vortex in the vicinity of a boundary.
    \label{cfmapping}}
  \end{center}
\end{figure}

\begin{figure}[t]
  \begin{center}
    \includegraphics[width=0.95\columnwidth]{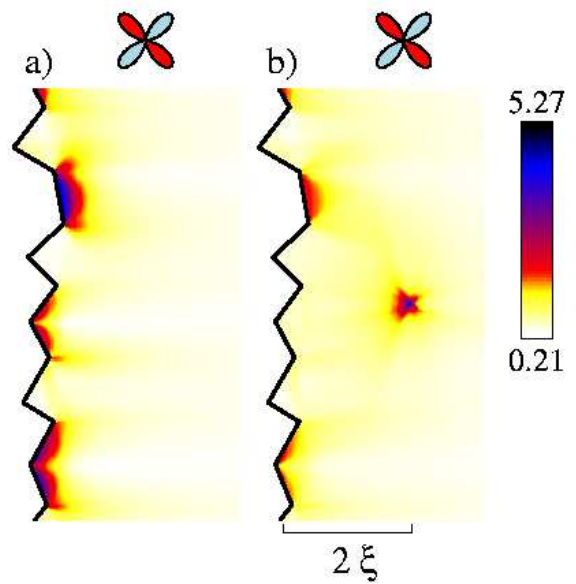}
  

    \caption{Zero-energy density of states of a $d_{x^2-y^2}$-wave 
    superconductor with a polygonal boundary geometry, which can
    be regarded as a boundary line with macroscopic roughness (faceting). 
    a) No vortex is present. 
   b) A single Abrikosov vortex is situated $2\xi$ away from 
   the averaged boundary line.  
    \label{Polygonal}}
  \end{center}
\end{figure}

We now examine the zero-energy local density of states of a $d$-wave
superconductor along a macroscopically rough boundary line. 
As an adequate model, we take a superconductor with a polygonal boundary geometry,
consisting of piecewise smooth facets, each of a length of the order of $\xi$.
In contrast to the model of a microscopically rough surface presented above,
the extension of the polygonal boundary geometry is taken into account completely.
For the more simple wedge-shaped boundary in section III, we obtained the
phase of the gap function in an analytical way by conformal mapping (cf. section II~B). 
For the more general case of a polygonal boundary, it becomes necessary to calculate
the Schwarz-Christoffel mapping \cite{Trefethen} numerically. 
We used an implementation by Driscoll\cite{Driscoll} to obtain 
the correct corresponding phase. This toolbox uses the analytically known 
solution of the phase distribution for a vortex sitting in the center of a disk
and conformally maps this solution onto the vortex close to the polygonal boundary,
as illustrated in Fig.~\ref{cfmapping}. For the disk, the lines of constant phase
are just radial lines, and the phase gradient corresponds to concentric circles,
fulfilling the boundary condition that the phase gradient is parallel to the
surface. The toolbox provides the mapping which maps the center of the
disk onto the vortex position and the boundary of the disk onto the polygonal
boundary. The resulting lines of constant phase and phase gradient after this
mapping are shown in  Fig.~\ref{cfmapping}, right panel.
This means, that within the approximation (\ref{gradandj}), the currents due to a
single Abrikosov vortex are forced to be parallel to each facet of the polygonal boundary.
Furthermore, the polygonal boundary allows for multiple reflections of quasiparticle 
trajectories.
Thus, the effects discussed in section~III can be found here as well.

In Fig. \ref{Polygonal}a) and b), the zero-energy local density of states of the $d$-wave 
superconductor with polygonal boundary structure is shown. The main direction of the boundary 
is oriented in such a way, that facets parallel to it exhibit maximum Andreev bound states.  
If we compare the local density of states shown in Fig. \ref{Polygonal}a), where no vortex is present, 
to the case with Abrikosov vortex presented in b), many of the effects already discussed above 
can be found again (cf. for example Fig. \ref{KeilVortex}).
The energy spectrum in the corner of a right-angled wedge is, independent of its 
orientation, the $d$-wave bulk spectrum. Thus, the nearly right-angled parts of the boundary exhibit
a very low zero-energy density of states, which is hardly affected by the presence of the vortex. 
The main effect of the vortex is again a strong suppression of zero-energy bound states in its shadow region. 
This is because the zero-energy spectral weight is effectively shifted towards higher energies.
Bound states at the boundary, which are further away, are reduced more moderately.

\section{Conclusions}

In this work, we investigated the influence of both the boundary 
geometry and a single pinned Abrikosov vortex on the zero-energy
local density of states along the boundary of a $d_{x^2-y^2}$-wave 
superconductor.
We found that a wedge-shaped boundary can induce a quasiparticle
spectrum in the corner of the wedge, which is completely different
to that farther away from the corner. The main mechanism for this
effect is the multiple reflection of quasiparticle trajectories. In the 
corner, this  can effectively be described as a mixing of bulk trajectories belonging
to different angles. Depending on the opening angle of the wedge and the orientation 
between $d$-wave and boundary geometry, this mixing can either lead to bulk spectra in
the corner (and thus to an absence of bound states) or to the presence of induced zero-energy bound states. 
In Fig.~\ref{AlpGamN}a), we have provided a map, showing 
for which combinations of opening angle and $d$-wave orientation bound states appear.
If the wedge is oriented in such a way, that the nodal direction of the $d$-wave is
parallel to the bisecting line, the zero-energy density of states in the corner strongly 
oscillates as a function of the opening angle of the wedge.
Although all results presented here have been obtained using a model gap function,
a fully selfconsistent calculation is not going to change the 
results qualitatively. The amplitude of the oscillations mentioned above, for example, 
should be somewhat reduced, especially for smaller opening angles. But in any case,
it should be possible to observe the oscillations at least for larger opening angles.   

Another example of an induced quasiparticle spectrum is a wedge with opening
angle $\pi/2$:  
The local density of states in the corner of a right-angled wedge is always given by a $d$-wave
bulk spectrum, even if the orientation allows the highest Andreev bound
states at both boundary lines. In addition to this geometrically induced prohibition 
of bound states, we presented the influence of a single pinned Abrikosov vortex on 
the zero-energy local density
of states at the boundary. Bound states at the boundary are locally suppressed, in particular
in a shadow-like region close to the vortex. However, this kind of suppression is due 
to a splitting 
of the high zero-energy peak of the bound states. Thus it is of a different physical
origin. The spectral weight is effectively shifted
towards higher energies, because the quasiparticles "see" a locally varying energy shift along
their trajectory, which stems from the locally varying flow field around the vortex.
Both effects can be observed independently in our results.    

We have studied the influence of two types of surface roughness on the vortex shadow
effect: microscopic, random scattering at the surface as well as a more macroscopic
faceting of the boundary. In the second case we have taken care to calculate the
flow field with the appropriate boundary conditions using the conformal mapping
technique. In both cases the vortex shadow effect clearly persists.

\acknowledgments

C.~Iniotakis is grateful to the German National Academic Foundation. 
S.~Graser is supported by the 'Graduiertenf\"orderungs\-programm des Landes
Baden-W\"urttemberg'.

\end{document}